# Network Function Virtualization based on FPGAs: A Framework for all-Programmable network devices

Christoforos Kachris, Georgios Sirakoulis
Electrical and Computer Engineering Department
Democritus University of Thrace
Xanthis, Greece
ckachris, gsirak @ ee.duth.gr

Dimitrios Soudris
Department of Electrical and Computer Engineer
National Technical University of Thrace
Athens, Greece
dsoudris @ microlab.ntua.gr

*Abstract*— Network Function Virtualization (NFV) refers to the use of commodity hardware resources as the basic platform to perform specialized network functions as opposed to specialized hardware devices. Currently, NFV is mainly implemented based on general purpose processors, or general purpose network processors. In this paper we propose the use of FPGAs as an ideal platform for NFV that can be used to provide both the flexibility of virtualizations and the high performance of the specialized hardware. We present the early attempts of using FPGA's dynamic reconfiguration in network processing applications to provide flexible network functions and we present the opportunities for an FPGA-based NFV platform.

## I. Introduction

Network and Telecom Operators are currently using several proprietary hardware appliances in order to meet the user's and the application's requirements. In order to deploy a new network service, they often need to acquire, configure and operate a new specialized hardware device that increases the cost, the power consumption and the complexity of the IT management. Furthermore, the specialized hardware-based appliances rapidly become obsolete, requiring much of the procure-design-integrate-deploy cycle to be repeated with little or no revenue benefit [1]. Moreover, as the lifecycle of the hardware becomes shorter as technology progress, the cost and the complexity of deploying these devices increases significantly.

"Network Functions Virtualisation aims to address these problems by leveraging standard Information Technology (IT) virtualisation technology to consolidate many network equipment types onto industry standard high volume servers, switches and storage, which could be located in Datacentres, Network Nodes and in the end user premises" [1].

The main benefits of using standard IT virtualisation technology to perform several network functions are the followings:

- Reduced equipment cost, by using commodity hardware resources instead of specialized hardware devices

- Decreased Time-to-Market by reducing the typical network operator cycle of acquiring-deploying-configuring-maintaining the specialized network hardware devices

- Higher flexibility by seamlessly upgrading the commodity hardware resources using updated software versions (often by 3$^{rd}$ parties)

- Higher flexibility and scalability depending on the applications requirements by using commodity hardware resources to scale up/down the required functions.

- Enables a wide variety of eco-systems and encourages openness. It opens the virtual appliance market to software entrants, Small and Medium Enterprises (SME) and academia, encouraging more innovation to bring new services streams quickly at much lower risk

However NFV in order to be feasible and can be deployed widely there are several challenges that need to be addressed:

- Achieving high performance virtualized network appliances in order to sustain the high data volume and the high processing complexity of the network functions

- Managing and orchestrating many virtual network appliances (particularly alongside legacy management systems) while ensuring security from attack and misconfiguration.

- Network Functions Virtualisation will only scale if all of the functions can be automated.

- Ensuring the appropriate level of resilience to hardware and software failures.

- Integrating multiple virtual appliances from different vendors. Network operators need to be able to "mix & match" hardware from different vendors, hypervisors from different vendors and virtual appliances from different vendors without incurring significant integration costs.

However, the most critical factor for the feasibility and the wide adoption of NFV is the ability to provide high performance packet processing and at the same time to provide high energy efficiency.

Taking all these in account, in this paper, we propose the use of FPGA-based Network Function Virtualization devices as an ideal platform that can provide both the flexibility of the virtualization and the high performance of the specialized hardware components.

In Section II we present in more detail the requirements and the architecture of NFV. In Section III we present the early attempts in the research literature to use FPGAs for network processing that can be easily reconfigured to meet new processing requirements. In Section IV we present our view on FPGA-based NFV devices as an ideal platform that can meet the requirement of NFV while at the same time it can support the high throughput requirements of future network traffic. Finally, in Section V we conclude on the proposed scheme and the future of FPGA-based NFV.

## II. Network Function Virtualization

Network Functions Virtualisation (NFV) aims to transform the way that network operators architect networks by utilizing standard IT virtualisation technology to consolidate many network equipment types onto industry standard high volume servers, switches and storage, which could be located in Datacentres and Network Nodes as illustrated in **Figure 1**. Specifically, it involves the implementation of network functions as software functions that can be hosted to a range of industry standard computing resources (i.e. commodity servers), and that can be moved to various locations in the network as required, without the need for installation of new equipment.

NFV aims to reduce the time to establish a new network service by just uploading the network functions of the service to the commodity hardware resources instead of purchasing new specialized hardware equipment. That way, network operators can very fast deploy new network services. Potential examples of NFV implementations could be:

- Switching elements: BNG, CG-NAT, routers.
- Tunnelling gateway elements: IPSec/SSL VPN gateways.
- Traffic analysis: DPI, QoE measurement.
- Service Assurance, SLA monitoring, Test and Diagnostics.
- Converged and network-wide functions: AAA servers, policy control and charging platforms.
- Application-level optimization: CDNs, Cache Servers, Load Balancers, Application Accelerators.
- Security functions: Firewalls, virus scanners, intrusion detection systems, spam protection.

However, the main weakness of this approach is the low throughput that the software-based network functions provide. The use of standard hardware resources (i.e. servers, switches) can reduce the overall cost, the time to market and also increase the flexibility; however the main problem is the latency and the throughput that the standard hardware can provide.

With the expected growth of the Internet and data center traffic in the next years, the main target of the system vendors is more powerful network devices that can sustain the increased traffic. Therefore, although that NFV based on standard hardware can provide the flexibility that is required by the telecom operators, in most of the cases it cannot sustain the high throughput requirements.

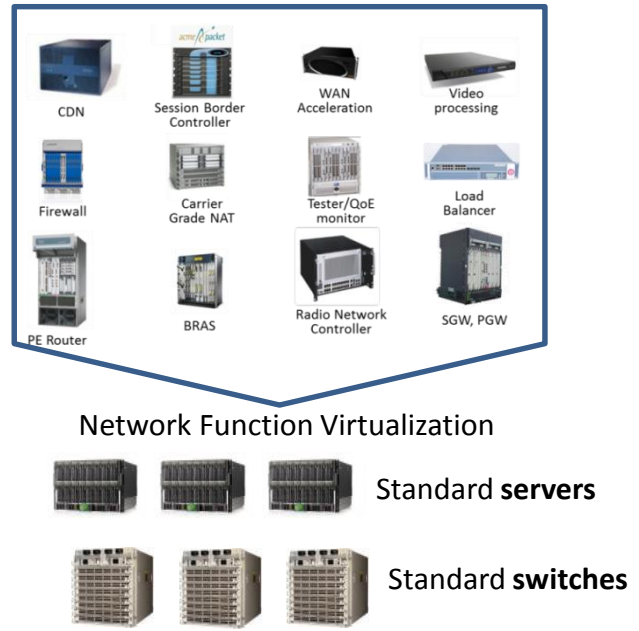

Figure 1. Network Function Virtualization

## III. Dynamic Reconfiguration and Network Processing

Programmable chips based on reconfigurable logic (FPGAs) have traditionally been used widely in the domain of network processing, both in the domain of wired and wireless communications. FPGAs can provide rapid prototyping to meet the Time-to-Market constraints and at the same time they can provide high throughput due to the customized hardware. Therefore, it has been shown that FPGAs can be utilized in the several cases of network processing applications. Table 1 shows some of the typical network processing applications that FPGAs have been used to speed up the processing of the network packets.

FPGAs can be programmed both statically and dynamically. Static reconfiguration means that the whole FPGA is stopped and configured with a new configuration file (bitstream). However, in most of the industrial network applications the FPGAs were mainly configured at design time

(static configuration), and when a new upgrade was available for the specific application.

Table 1 FPGA implementations in network applications

| Application | Reference |
|---|---|
| Firewall | [3] |
| Intrusion Detection | [4][5][6] |
| Content-based switching | [7] |
| Packet Parsing | [8] |
| IP Lookup | [9][10] |
| Deep packet inspection | [11][12] |
| Worm/Virus | [13] |

On the other hand, dynamic (partial) reconfiguration refers to the option to configure part of the FPGA, while the rest of the FPGA is operating without any disruption. In the past there were several research efforts to utilize the dynamic reconfiguration of the FPGA in order to design reconfigurable systems that can adapt to the processing requirements.

In [14] and [15], a novel reconfigurable network processing has been presented that can dynamically change the number and the type of hardware accelerator units in order to meet the network traffic fluctuations. Specifically, the network processing platform can dynamically reconfigure the number of compressions, and encryption hardware acceleration units to meet the processing requirements of the network traffic. Furthermore, the proposed platform can change the number of hardware acceleration units that are based tightly coupled and loose coupled (through a bus) to the processor based on the average size of the network packets. The proposed reconfigurable platform could be used for example in an edge router where the number of flows that needs encryption (i.e. VPN connection) and compression (i.e. wireless connections) change dynamically over time.

The utilization of FPGA for network processing application has been also proposed by Lockwood et al. in [16]. The proposed platform allows reprogrammable hardware modules to by dynamically installed into a router or firewall through the use of full or partial reprogramming of an FPGA. The applications that have been developed for this platform include Internet packet routing, data queuing and application level content modification. These notions have been also implemented in the NetFPGA platform that allows the design and development of network processing devices based on an FPGA-base open platform [17].

In [18], a reconfigurable network co-processor platform was presented called DynaCore. In that paper an FPGA-based platform was presented that could accommodate hardware accelerator units. The platform included a dispatcher that was used to forward the incoming packets to the hardware acceleration units. The system was consisted only of hardware acceleration units without presenting a connection of the hardware units with the general-purpose processing elements used for the remaining header processing.

In [19], a reconfigurable platform was introduced targeting mainly active networks. This system consists of software and hardware parts. The software part was a set of kernel and user space modules running on a Linux PC. The hardware part was consisting of an FPGA device that was used to load the required processing modules. When a packet was received, it was checked whether it was a passive or an active packet. For active packets, the system checks whether the required hardware for this application is already present in the device. Otherwise, it can request the bitstream for the specific active packet. When a new bitstream is received for an active packet, the bitstream is authenticated, decrypted and checked for integrity and then is used for the configuration of the device.

In [20], the design and analysis of a network processor using accelerators in reconfigurable logic was presented. In that paper, two different approaches were presented. In the first case, each task is mapped to a general-purpose accelerator. In the second case, different accelerators are used for different tasks that can be dynamically reconfigured on the device. The paper showed that the use of reconfigurable modules can improve the execution time by about 20 times. The system has been evaluated in three applications: tree-lookup, pattern matching, and network intrusion detection.

## IV. FPGA-based Network Function Virtualization

In order to ease the rapid development of network applications using the flexibility of the software, several chip vendors have proposed programmable platforms that are based on specialized network processors.

EZchip Technologies, provider of one of the highest-performance packet forwarding engines is currently shifting from its traditional data plane (NP) family to the new NPS, which turns to a multithreaded CPU running Linux and programmed in C [21].

Marvell offers two all-inclusive data plane software suites for its Xelerated processor [22]. Both the Metro Ethernet Application and the Unified Fiber Access Application consist of an application package running on the NPU, and a control-plane API running on host CPU. The latter API is based on a hardware adaptation layer that includes boot scripts, configuration modules, and predefined messages to access the forwarding plane from control plane. This may represent an easy way for an OEM customer to bring a product to market, but modifications are only made through Marvell-supplied source code – the processor itself uses assembly code.

Cavium has opted for a more generic Software Development Kit for its Octeon family [23]. The SDK includes a Gnu toolchain, simulator, Cavium's own ViewZilla for graphical analysis, and a regular-expression pattern compiler for deep packet analysis. Vertical software toolkits provide C-based routines for SSL, TCP, IPsec, and similar common Layer 3-5 functions. A general-purpose control plane CPU with some datapath elements can be highly flexible, but is far from optimal for packet processing.

Recently, Xilinx has released the SDNet approach [24]. According to Xilinx, SDNet allows for the creation of 'Softly' Defined Networks. 'Softly' Defined Networks supports Software Defined Networking (SDN) functionality while also allow differentiation through software programmable data plane hardware. The programmable hardware can dynamically collaborates with control plane SW while addressing the performance, flexibility, and security challenges of modern content-oriented networking.

Although that SDNet is mainly proposed for Software Defined Networking, a similar approach could also be used for the utilization of FPGAs in Network Function Virtualization (NFV). The FPGAs could be used as the main programmable platform for the realization of NFV by configuring the FPGA with the required hardware accelerators. For example, in the case of Deep Packet Inspection (DPI) or Firewall applications, the FPGA could be configured to host pattern string matching accelerators; In the case of Load Balancing units it could configured with the right load balancing accelerators, etc.

By using FPGAs as the main hardware, NFV could achieve the flexibility and the fast deployment of the software-solutions and the high throughput of the specialized hardware devices. A possible framework for the efficient deployment of FPGA-based NFV is depicted in Figure 2.

As shown in this figure, the FPGAs could be configured with the hardware components that are required for the specific application from a pool (library) of $3^{rd}$ party hardware components. One of the advantages of the proposed scheme is that it will allow the fostering of a new ecosystem in the domain of network devices including FPGA vendors, IP block vendors, and system integrators. By exposing the underlying hardware to the user s (i.e. IP vendors), there is the advantage of higher flexibility, higher competiveness and higher forms of innovation. Furthermore, an FPGA-based NFV platform could host hardware accelerators from different vendors. For example, as shown in the figure, the Encryption and the Compression modules could be provided by Vendor A while the content-based switching module could be provided by Vendor B. In order to facilitate the inter-operability of the hardware components from different vendors, a standard interface between the modules could be defined (NFV-IF in the figure). For example, the communication between the hardware accelerators could be based on System Packet Interface (SPI) 4.2 [26] or any other kind of standard interface for the efficient exchange of packets between the hardware modules. On the other hand, a standard common interface could be used for the configuration of the $3^{rd}$ party hardware modules from the NFV controller. The NFV controller can be used to load the components from the library, program the PFGA and then configure the modules based on the application's requirements through the NFV-Configuration Interface (NFV-Config-IF).

The proposed scheme could be utilized complementary to the notion of the SDN/OpenFlow [25]. While SDN is mainly focused on the programmability of the control plane, NFV is mainly focused on the programmability of the data plane. Therefore, the proposed FPGA-based NFV devices could be used by the seamlessly co-existence of the SDN for the control plane and the NFV for the programmable data plane driving towards a fully programmable network devices.

As a possible reference example we describe the operation of an edge router implemented using the FPGA-based NFV platform. Edge routers are routers that provide authenticated access to faster, more efficient backbone and core networks. Edge routers need to support multi-service functions to manage different types of network traffic. For example, they may need to support encryption of the traffic or packet compression. Using the FPGA-based NFV framework, the NFV controller could load from the $3^{rd}$ party library the required modules to support network encryption and network compression. Due to traffic fluctuations, if the number of network flows that need encryption increases and the number of flows that need compression reduces, then the NFV-controller could program the FPGA with additional modules for encryption (or with more efficient modules for encryption that allocate more area). That way, the FPGA-based NFV could adapt its resources based on the network traffic fluctuations.

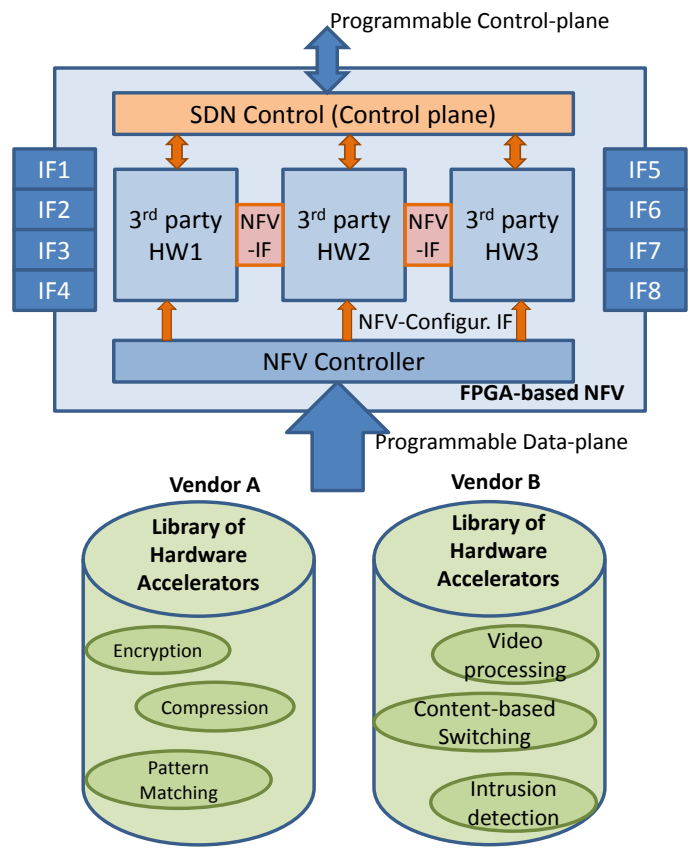

Figure 2. FPGA-based NFV

However, the use of FPGA could be justified only in cases that the network operator is willing to sacrifice the high performance of the dedicated hardware components (i.e. based on ASIC) with the lower cost and flexibility that the FPGA-based NFV could provide. Figure 3 shows where FPGA-based NFV device could be located depending on the application's requirements and the network operator's requirements, respectively. In applications where the network operators are mostly interested about the flexibility, the low cost and the fast deployment of the services, the GPP-NFV platforms could be used. In applications where the most important factor is the performance (i.e. high throughput in core networks) and the low power consumption, the current application-specific devices will still be used. However, in applications where we need both a high degree of flexibility and also a high degree of performance, FPGA-based NFV platform could be used as a promising platform.

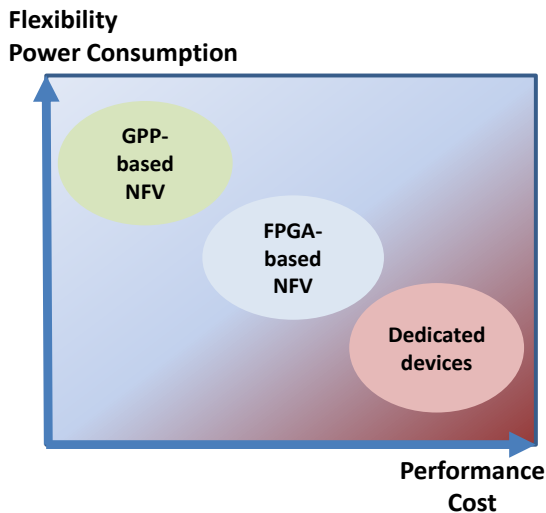

Figure 3. Design Space Exploration of NFV solutions

## v. Conclusions

In this paper, we introduce an alternative way to implement NFV to target the future requirement of the network operators based on FPGAs. FPGA-based NFV can provide the flexibility of the NFV based on general purpose processors (GPP's) while at the same time providing the necessary throughput that the GPP cannot sustain.

The proposed scheme could be used to foster further innovation in the domain of hardware IP ecosystem. FPGA-based NFV could be used to expose the hardware resources of the network devices to the IP vendors, in the same way that SDN/OpenFlow has been used to expose the resources of the routers to the control plane users. By efficiently combining the SDN and the NFV in the FPGA platform we can achieve the design of all programmable network devices fostering the innovation to a new ecosystem for IP vendors in network applications.

## *References*

[1] "Network Functions Virtualisation— Introductory White Paper". ETSI. 22 October 2012. Retrieved 20 June 2013.

[2] "Cisco Visual Networking Index: Forecast and Methodology, 2013-2017", Cisco White paper, May 2013

[3] A. Wicaksana, A. Sasongko, "Fast and reconfigurable packet classification engine in FPGA-based firewall", Electrical Engineering and Informatics (ICEEI), 2011 International Conference on, July 2011

[4] C. R Clark, C. D. Ulmer, D. E. Schimmel, "An FPGA-based network intrusion detection system with on-chip network interfaces", IEEE Transactions on Information Forensics and Security, (Volume:3 , Issue: 1 ), March 2008

[5] I. Sourdis, D. Pnevmatikatos, S. Vassiliadis, "Scalable Multi-Gigabit Pattern Matching for Packet Inspection", IEEE Transactions on Very Large Scale Integration (VLSI) Systems, Special Section on Configurable Computing Design-II: Hardware Level Reconfiguration, Vol. 16, Issue 2, pp. 156-166, February 2008

[6] A. Mitra, W. Najjr, and L. Bhuyan, "Compiling PCRE to FPGA for accelerating SNORT IDS". Proceedings of the 3rd ACM/IEEE Symposium on Architecture for Networking and Communications Systems, Aug. 2007, pp. 127-136.

[7] C. Kachris, S. Vassiliadis, Design of a Web Switch in a Reconfigurable Platform, ACM/IEEE Symposium on Architectures for Network and Communication Systems (ANCS 2006), San Jose, December 2006

[8] M. Attig, G. Brebner, "400 Gb/s Programmable Packet Parsing on a Single FPGA", Seventh ACM/IEEE Symposium on Architectures for Networking and Communications Systems (ANCS), October 2011

[9] Weirong Jiang, V. K. Prasanna: Data Structure Optimization for Power-Efficient IP Lookup Architectures. IEEE Trans. Computers 62(11): 2169-2182 (2013)

[10] T. Ganegedara, V. K. Prasanna: "A high-performance IPV6 lookup engine on FPGA", FPL 2013: 1-4

[11] Yi-Hua E. Yang, V. K. Prasanna, "Robust and Scalable String Pattern Matching for Deep Packet Inspection on Multicore Processors", IEEE Trans. Parallel Distrib. Syst. 24(11): 2283-2292 (2013)

[12] S. Dharmapurikar and J. W. Lockwood, "Deep packet inspection using parallel bloom filters," Micro, IEEE, vol. 24, no. 1, pp. 52–61, 2004

[13] J. W. Lockwood, J. Moscola, D. Reddick, M. Kulig, and T. Brooks, "Application of hardware accelerated extensible network nodes for internet worm and virus protection," pp. 44–57, 2003

[14] C. Kachris, S. Wong, S. Vassiliadis, "Design and Performance Evaluation of an Adaptive FPGA for network applications", Microelectronics Journal, 2008, 40 (7), pp.1103-1110

[15] C. Kachris, "Reconfigurable Network Processing Platforms", Ph.D. Thesis, Delft University of Technology, 2007

[16] J.W. Lockwood, "An open platform for development of network processing modules in reprogrammable hardware", pp. WB-19, January 2001

[17] J. W. Lockwood, et al., "Netfpga–an open platform for gigabit-rate network switching and routing," in Microelectronic Systems Education, 2007. MSE '07. IEEE International Conference on, 2007, pp. 160–161

[18] J. Foag, R. Koch, Architecture conception of a reconfigurable network coprocessor platform (DynaCore) for flexible task offloading, ANCHOR 2004, Munchen, 2004, pp. 32–38

[19] N.G. Bartzoudis, et al., Reconfigurable computing and active networks, Engineering of Reconfigurable Systems and Algorithms (2003) 280–283

[20] G. Memik, S.O. Memik, W.H. Mangione-Smith, Design and analysis of a layer seven network processor accelerator using reconfigurable logic, in: IEEE Symposium on Field-Programmable Custom Computing Machines (FCCM'02), Napa, CA, USA, April 2002.

[21] "NP-5, 240Gbps NPU for Carrier Ethernet applications", Product Brief, EZChip, 2014

[22] "Marvell Xelerated HX300 Family of Network Processors", Datasheet, Marvel Inc, 2013

[23] "Octeon III CN70xx/71xx Network Processor Families", Product Brief, Cavium Inc., 2014

[24] L. Wirbel, "Xilinx SDNet: A new Way to Specify Network Hardware", The Linley Group, White paper, March 2014

[25] N. McKeown et al, "OpenFlow: Enabling Innovation in Campus Networks", ACM SIGCOMM Computer Communication Review, vol. 38, no.2 April 2008

[26] "System Packet Interface Level 2 (SPI 4.2)", Optical Internetworking Forum, White paper, October 15, 2013, www.oiforum.com